\documentclass[runningheads]{llncs}
\usepackage[T1]{fontenc}
\usepackage{amssymb}
\usepackage{amsmath}
\usepackage{siunitx}
\usepackage{array}
\usepackage[caption=false,font=normalsize,labelfont=sf,textfont=sf]{subfig}
\usepackage{textcomp}
\usepackage{stfloats}
\usepackage{booktabs}
\usepackage[hyphens]{url}
\usepackage[hidelinks]{hyperref}
\usepackage{verbatim}
\usepackage{graphicx}
\usepackage{cite}
\usepackage{xcolor}
\usepackage[ruled,linesnumbered]{algorithm2e}
\usepackage{multirow}
\usepackage{orcidlink}
\usepackage[misc,geometry]{ifsym}
\usepackage{wrapfig}
\usepackage{caption}
\usepackage{float}

\begin{document}

\title{A Simulation Framework for Electromagnetic Signal Injection Attacks on Image Sensors}
\titlerunning{A Simulation Framework for ESIA on Image Sensors}

\author{Youqian Zhang\inst{1}\orcidlink{0000-0003-0907-7998}\and
MK Michael Cheung\inst{1}\orcidlink{0009-0005-3591-9089}\and
Chunxi Yang \inst{2}\orcidlink{0009-0001-5435-6083}\and
Xinwei Zhai\inst{2}\orcidlink{0009-0005-3685-9591} \and
Zitong Shen\inst{1}\orcidlink{0009-0008-4950-0627} \and
Xinyu Ji\inst{1}\orcidlink{0000-0003-1109-9766} \and
Eugene Yujun Fu\inst{2}\orcidlink{0000-0003-1048-1904}\Letter \and
Sze Yiu Chau\inst{3}\orcidlink{0000-0001-9300-0808} \and \\
Xiapu Luo\inst{1}\orcidlink{0000-0002-9082-3208} 
}

\authorrunning{Y. Zhang et al.}

\institute{
The Hong Kong Polytechnic University, Hong Kong \and
The Education University of Hong Kong, Hong Kong \and
Simon Fraser University, Canada \\
\Letter\email{eugenefu@eduhk.hk} 
}

\maketitle
\begin{abstract}
Image sensors are fundamental to many intelligent systems, allowing visual perception and AI-driven decision-making. However, their integrity can be compromised by electromagnetic signal injection attacks (ESIA), which manipulate captured images without modifying sensor hardware or software. Despite the growing threat, system-level understanding of the attacks, as well as the development of defenses, remains limited, in part because collecting adversarial data is often complex and requires specialized attack setups. To address this challenge, we model ESIA and develop a simulation framework for generating synthetic adversarial images. Our analysis shows that these synthetic images are statistically indistinguishable from those produced by real attacks. The proposed framework enables faster vulnerability evaluation of computer vision (CV) algorithms, without the need for dedicated attack hardware. We also present a pilot study showing that the robustness of the algorithms can be improved by adversarial training, demonstrating a practical and scalable path toward mitigating ESIA threats.
\end{abstract}


\section{Introduction}
\label{sec:introduction}
The rapid advancement of artificial intelligence (AI) has transformed numerous devices and applications into intelligent systems.
These systems can perceive and interact with the physical world through various sensors.
Image sensors are one of the widely adopted, particularly in many safety- and security-critical applications of intelligent vision systems such as surveillance, autonomous vehicles, and industrial quality control, where image data are captured and processed to make critical decisions. 
For the AI to function reliably, the integrity of captured images is crucial. 
However, prior work has demonstrated that attackers can mislead AI by manipulating images using adversarial samples~\cite{guesmi2023physical}, or physical interference such as laser/light~\cite{petit2015remote,yan2016can,fu2021remote,wang2021can,kohler2021they,yan2022rolling,man2024remote} and ultrasound~\cite{ji2021poltergeist,cheng2023adversarial,zhu2023tpatch}.

\begin{figure*}[t]
\centering
\begin{minipage}[h]{0.23\textwidth}
  \vspace{0pt}
  \centering
  \includegraphics[width=\linewidth]{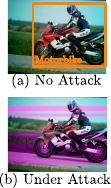}
  \captionof{figure}{Purple strips by ESIA.}
  \label{fig:practical_scenario}
\end{minipage}
\hfill
\begin{minipage}[h]{0.62\textwidth}
  \vspace{0pt}
  \centering
  \captionof{table}{Impacts of ESIA on Different CV Tasks}
  \label{tab:esia_on_image_sensor_previous_work}
  \footnotesize
  \setlength{\tabcolsep}{4pt}
  \begin{tabular}{@{}llll@{}}
  \toprule
  \textbf{Work} & \textbf{CV Task} & \textbf{Metric} & \textbf{Range} \\
  \midrule
  \multirow{2}{*}{\cite{jiang23glitchhiker}}
    & Object Detection  & ASR & 81.0\% -- 99.3\% \\
    & Face Recognition  & ASR & 69.8\% -- 84.0\% \\
  \midrule
  \multirow{1}{*}{\cite{zhang2024esia}}
    & Object Detection  & ASR &  5.32\% -- 86.7\% \\
  \midrule
  \multirow{1}{*}{\cite{lu2026phantom}}
    & VLA  & ASR &  0.8\% -- 84\% \\
  \midrule
  \multirow{1}{*}{\cite{zhang2025rainbow}}
    & Object Detection  & DEG &  $-$23\% -- $-$4\% \\
  \midrule
  \multirow{2}{*}{\cite{liao2025your}}
    & Object Detection  & DEG & $-$65\% -- $-$11\% \\
    & Segmentation      & DEG & $-$14\% -- $-$2\% \\
  \bottomrule
  \end{tabular}
  \smallskip
  \begin{minipage}{\linewidth}
    \footnotesize
    ASR = Attack Success Rate;\quad
    DEG = Performance Degradation;\quad
    VLA = Vision Language Action Model
  \end{minipage}
\end{minipage}
\vspace{-1em}
\end{figure*}

There is an emerging and novel attack vector using electromagnetic waves to disrupt image transmissions, resulting in image distortions~\cite{kohler2022signal,jiang23glitchhiker,zhang2024esia,liao2025your,lu2026phantom,liu2025magshadow,ren2025ghostshot}. 
In this work, we refer to such threats as \textit{electromagnetic signal injection attacks (ESIA)}.
ESIA can introduce visible artifacts into captured images, such as the purple strips shown in Fig.~\ref{fig:practical_scenario}(b). More critically, these artifacts can mislead downstream computer vision (CV) tasks, leading to failures such as misidentification in surveillance systems or incorrect object recognition in robotic platforms. 
For instance, an attacker may manipulate the image sensor of an autonomous vehicle to mask a motorcycle in the captured frames (as shown in Fig.~\ref{fig:practical_scenario}), potentially causing the vehicle to miss the motorcycle and increasing the risk of a collision~\cite{liao2025your}.

Despite increasing awareness of ESIA, many CV algorithms remain underexplored because studying these attacks in practice is difficult. 
ESIA experiments require specialized hardware to generate various electromagnetic interference, and the lack of adversarial images produced by ESIA makes it difficult to develop and evaluate effective defenses.
Moreover, real-world testing is costly and time-consuming, as it depends on access to target devices and controlled environments.
To address these challenges, we make the following contributions:
\begin{itemize}
    \item \textbf{Modeling and Simulation Framework:} We propose a generalized system model and a row-drop-based simulation method for generating synthetic adversarial images, enabling efficient and scalable evaluation without physical attacks (Section~\ref{sec:modeling_and_principle}).
    
    \item \textbf{Evaluation of Synthetic Adversarial Images:} We show that the synthetic patterns closely match real-world attacks at three levels: raw image, reconstructed image, and AI-model levels, supported by statistical analysis (Section~\ref{sec:evaluation}).

    \item \textbf{Vulnerability Assessment and Robustness Enhancement:} Using the simulation framework, we identify additional vulnerable CV tasks and algorithms and show that fine-tuning with synthetic adversarial images could improve robustness against ESIA (Section~\ref{sec:case_study}).
\end{itemize}


\section{Background and Related Work}
\label{sec:background}



In electronic devices, cables (or traces) act as communication pathways, connecting different components. 
These cables, essentially made of conductive metals like copper, possess inherent antenna-like properties, meaning that they can capture environmental electromagnetic waves~\cite{wilson2010radiation,paul2022introduction}. 
Such antenna-like behavior creates an attack surface where attackers can inject electromagnetic signals.
The injected signals will be superimposed with original signals that are transmitted in the cables, changing signal waveforms. 
Regarding digital signals represented as binary 0s and 1s, injected signals can distort the waveform and cause a voltage corresponding to one bit to be interpreted as the other. 
As a result, ESIAs can introduce bit errors in data transmission, a phenomenon validated by many prior studies~\cite{zhang2024virtual,jang2023paralyzing,jiang23glitchhiker,dayanikli2022wireless,kohler2023brokenwire}.

Specifically in this work, we focus on ESIA that primarily targets Complementary Metal-Oxide-Semiconductor (CMOS) image sensors, which is the most widely used type\footnote{This work does not consider Charge-Coupled Device (CCD) image sensors or the effects of the cited attacks~\cite{kohler2022signal,liu2025magshadow,ren2025ghostshot}, as CCDs are less common now due to their higher cost compared to CMOS sensors~\cite{insights2022cmos}.}.
Unless otherwise stated, ``image sensors'' hereafter refer to ``CMOS image sensors''.
The impact of ESIA on the image sensors includes color strips, as mentioned previously. 
Such color strips have already demonstrated significant vulnerabilities across several computer vision tasks, including object detection~\cite{jiang23glitchhiker,zhang2024esia,liao2025your,zhang2025rainbow}, face recognition~\cite{jiang23glitchhiker}, segmentation~\cite{liao2025your}, and vision language action (VLA)~\cite{lu2026phantom} models, as summarized in Table~\ref{tab:esia_on_image_sensor_previous_work}.
The attacks can have a substantial impact, with attack success rates (or ASR, defined as the proportion of images that lead to failures in the target task) or performance degradation (DEG, measured as the absolute difference or percentage drop in evaluation metrics between the original model and the model under attack) reaching alarming levels. 
For example, the initial work by Jiang et al.~\cite{jiang23glitchhiker} has shown ASR between 81\% and 99.3\% on object detection tasks, possibly leading the system to detect something that does not exist or to hide an object.

Although many studies have presented successful attacks on the image sensors, few defenses have been demonstrated in prior work. 
Conventional approaches, such as electromagnetic shielding, can reduce external interference, yet unavoidable openings, including optical apertures~\cite{backstrom2004susceptibility} and cable penetrations~\cite{schulz1968elf}, may degrade shielding effectiveness. Jiang et al.,~\cite{jiang23glitchhiker} proposed a detection method that modifies protocols to identify disrupted data packets\footnote{It was reported that there is a dramatic difference between images of ``no attack'' and ``under attack'', but no quantified detection rate was reported.}, while Deng et al.,~\cite{kang25anti} sought to preserve downstream AI performance by intentionally discarding part of the image\footnote{The method preserves at least around 80\% of a model's original performance.}. However, methods for mitigating attack impacts remain underexplored (especially on the AI level), which we believe is partly due to the limited availability of adversarial data. 
By grounding our simulation framework in the underlying attack mechanism, our method can help address this gap and accelerate defense development.
\section{Modeling and A Simulation Framework}
\label{sec:modeling_and_principle}
This section begins by abstracting a high-level system model commonly found in different image sensor systems and presenting an attacker model, stating their capabilities and limitations.
Based on the models, we further elaborate on the attack principles and the method that generates synthetic adversarial images.

\subsection {System Model}
\label{sec:system_model}
\begin{figure*}[t]
\centering
\includegraphics[width=0.90\textwidth]{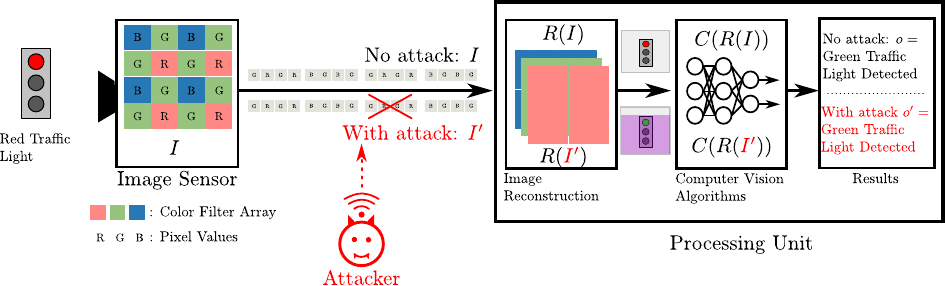}
\caption{An image sensor system consists of an image sensor and a processing unit.}
\label{fig:system_model}
\vspace{-1em}
\end{figure*}

An image sensor system, as shown in Fig.~\ref{fig:system_model}, consists of an image sensor and a processing unit. The sensor captures light from the physical world and converts it into electrical signals using photodiodes, each corresponding to one pixel. A color filter array (CFA) is placed over the photodiodes so that each pixel records only one color channel. A common CFA is the Bayer filter array~\cite{bayer1976color}, which arranges red, green, and blue filters in a fixed pattern~\cite{maschal2010review}. The resulting pixel values form a raw image \(I\).

The raw image is transmitted row by row from the sensor to the processing unit, with each row sent as a data packet. 
Among image sensors that are tested in this study, if a packet is corrupted, the receiver drops it and replaces it with the subsequent packet~\cite{jiang23glitchhiker}. The processing unit then reconstructs the image, primarily through demosaicing, which estimates missing color values from neighboring pixels, along with other steps such as noise reduction and white balancing. We denote the overall reconstruction function as \(R(\cdot)\), so that \(R(I)\) is the reconstructed image.

The reconstructed image is then processed by computer vision algorithms, such as object detection, segmentation, or face recognition, denoted by \(C(\cdot)\). The final system output is therefore
\(
o = C(R(I)).
\)

\subsection{Attacker Model}
\label{sec:attacker_model}

The attacker's objective is to cause the computer vision algorithm to make incorrect decisions by using electromagnetic waves to interfere with the transmission of the raw image.
We represent such a raw image under the attack as $I^{\prime}$. 
We consider that the attacker cannot tamper with the image reconstruction function $R(\cdot)$ or the computer vision algorithms $C(\cdot)$.
The attacker possesses specialized equipment capable of generating arbitrary electromagnetic waves, allowing the waveforms to be fine-tuned for effective and efficient attacks; the attacker radiates the adversarial electromagnetic waves through antennas.
In our model, each attack is characterized by five physical-layer parameters, grounded in prior studies~\cite{jiang23glitchhiker,yan2020sok}, that determine how effectively the electromagnetic signal couples into the victim system. 
These parameters fall into three categories: (1) signal properties, namely frequency (Freq.) and radiated power (Pow.); (2) temporal behavior, namely attack duration (Dur.); and (3) spatial configuration, namely the distance (Dist.) and angular alignment (Ang.) between the transmitting antenna and the target hardware.



\subsection{Attack Principle: How ESIA Produces Color Strip Artifacts?}
\label{sec:attack_principle}

Recall that a cable is used to connect the image sensor and the processing unit. 
As discussed in Section~\ref{sec:background}, this cable presents a potential attack surface: while designed for data transmission, it also acts as an antenna, susceptible to external electromagnetic interference.
The attacker can exploit this antenna property by injecting arbitrary electromagnetic waves into the cable, causing bit flips and corrupting the data packet transmission between the image sensor and the processing unit.
As mentioned in the system model, the corrupted packet is discarded and the missing information is replaced by the subsequent packet.
Such a replacement will disrupt the demosaicing process, leading to color strips in the reconstructed image.

\begin{figure}[t]
\centering
\includegraphics[width=0.8\textwidth]{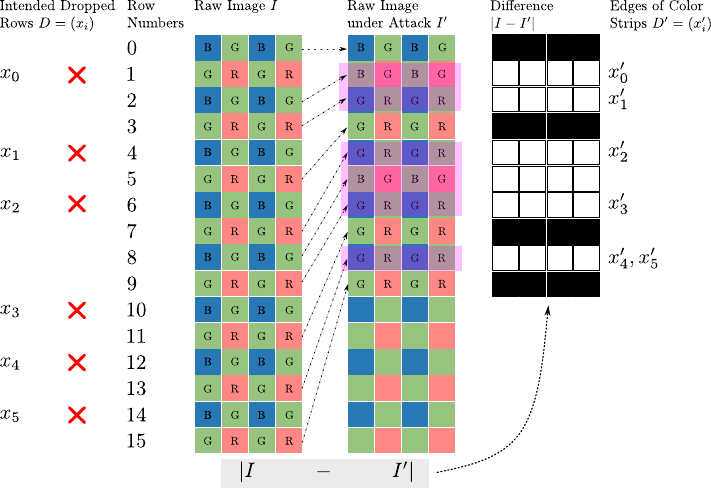}
\caption{Rows are dropped from a raw image $I$, leading to the raw image under the attack $I^{\prime}$. The empty places in $I^{\prime}$ will be filled by pixels from the next frame.}
\label{fig:killed_rows}
\vspace{-1em}
\end{figure}

\begin{figure}[t]
\centering
\includegraphics[width=1\textwidth]{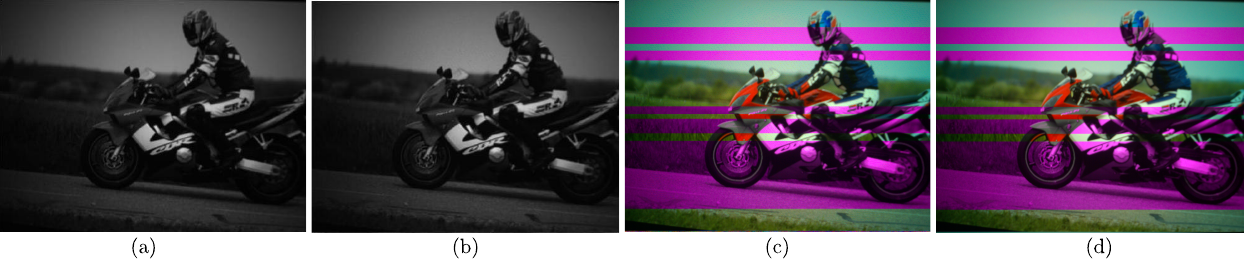}
\caption{(a) is the raw image without attack, i.e., $I$. (b) is the raw image under attack $I^{\prime}$. (c) is the reconstructed image of the raw image under attack $R(I^{\prime})$, showing multiple purple strips. (d) is the synthetic adversarial image $R(I^{s})$, which replicates the color strips the same as (c).}
\label{fig:bayer_vs_rgb}
\vspace{-1.2em}
\end{figure}

For example, Fig.~\ref{fig:killed_rows} illustrates how the color strips appear in an image. Let's assume an attack causes the loss of the 1st row in the ``Raw Image $I$'', which contains green (G) values and red (R) values, i.e., GRGR.
The processing unit fills the gap using information from the 2nd row of pixels, i.e., BGBG, where B represents blue.
This can result in incorrect color replacements during demosaicing because, as shown in ``Raw Image under Attack $I^{\prime}$'',  missing green values are replaced with B, and red with green G.
Similarly, in the 2nd row, its values are replaced by the 3rd row, thus blue is replaced with G, and green is replaced with R.
Such an incorrect color transformation will continue until another row of pixels is lost, for example, when the 4th row is lost.
As a result, a color strip is formed by causing two rows of pixels to drop in the reconstructed image.
With more rows being dropped, more strips appear. 
After the reconstruction, multiple color strips appear in the image, e.g., Fig.~\ref{fig:bayer_vs_rgb}(c).

\subsection{Simulation Method}
\label{sec:simulation_method}

In the following part, we introduce how to simulate adversarial color strips that can resemble real attacks' impacts.

\subsubsection{Characterizing Color Strips.}
To simulate the attack effects, it's essential to identify the rows of pixels targeted for removal, which mimics data loss during transmission caused by the attack.
The selection strategy for these rows depends on the desired impact on the image, such as introducing specific color strips to disrupt key image features.
This work focuses on a general simulation method, leaving the specific row selection method open for customization based on the application.

For a raw image $I$, we define a tuple $D$ including $m$ distinct indices of intended dropped rows, which are denoted as $x_{i}$, and the elements in $D$ are arranged in ascending order:
\(
    D = (x_{0},...,x_{m-1}).
\)
As discussed previously, a color strip is paired up with two dropped rows, and hence, the total number of strips is ${n = \lceil \frac{m}{2} \rceil}$.
Note that if two consecutive rows are dropped, there will not be any incorrect color transformation. 
Since this work focuses on generating the adversarial color strips, dropping rows consecutively is not considered in this work.

Regarding the positions of the color strips in an image, as demonstrated in Fig.~\ref{fig:killed_rows}, the $0^{th}$ strip starts at $x_{0}^{\prime}$ and ends at $x_{1}^{\prime}$, and the $1^{st}$ strip from $x_{2}^{\prime}$ to $x_{3}^{\prime}$, and so on.
In general, a color strip starts at $x_{i}^{\prime}$, where $i$ is an even number, and ends at $x_{i+1}^{\prime}$.
The height of the strip is $x_{i+1}^{\prime} - x_{i}^{\prime} + 1$, while the width of the strip is constant, which is the same as the width of the image.
It is essential to note that a special case is when ${i = m - 1}$ is even, and the strip will start from $x_{m-1}^{\prime}$ until the end of the image.

From Fig.~\ref{fig:killed_rows}, it can be observed that ${x_{i} \neq x_{i}^{\prime}}, i \in Z^{+}$, while $x_{0} = x_{0}^{\prime}$.
We can deduce the relationship between $x_{i}$ and $x_{i}^{\prime}$ so as to know where the strips will appear:
\begin{equation}
    x_{i}^{\prime} = 
    \begin{cases}
        x_{i}, & \text{if}~i = 0 \\
        x_{i}  - 2 \times \lfloor \frac{i + 1}{2} \rfloor, & \text{otherwise}
    \end{cases}
\label{eq:position_of_strips}
\end{equation}

While removing rows reduces image height by $m$, the processing unit will maintain the original dimensions for easier processing and analysis. 
As explained in our system model, this is achieved through padding, and for generality, we define the padding process as a function $P(I, m)$, meaning padding $m$ rows to the bottom of $I$.
For example, one approach involves appending the first $m$ rows of the original data to the bottom of the manipulated image. 
This simulates the scenario where the attack causes data from subsequent frames to shift upwards, filling the gap created by the dropped rows. 

The simulation process is summarized in Algorithm~\ref{algo:simulation}. 
It first sorts the dropped row indices in descending order allowing the removal to happen from the bottom upwards.
The loop iterates through each dropped row index and removes the corresponding row in a raw image.
Finally, the manipulated data is obtained by padding missing rows.
We denote the synthetic adversarial image as $I^{s}$, and the reconstructed image that will be further processed by computer vision algorithms is $R(I^{s})$.

\begin{algorithm}[H]
\scriptsize
\caption{Simulation Method}
\label{algo:simulation}
\SetKwInOut{Input}{input}\SetKwInOut{Output}{output}
\SetKwFunction{DescendingSort}{DescendingSort}
\SetKwFunction{Append}{Concatenate}
\SetKwFunction{Padding}{P}
\SetAlgoLined
\Input{A raw image $I$, a tuple $D$ of indices of rows to be dropped}
\Output{A raw image $I^{s}$ under synthetic attack impacts.}
$D \leftarrow$ \DescendingSort{$D$} \tcp{Sort dropped indices in a descending order.}
\For{$x_{i}$ in $D$}{
    $I \leftarrow$ \Append{$I[0:x_{i}-1,:]$, $I[x_{i}+1:, :]$} \tcp{drop $x_{i}$-th row.}
}
$I^{s} \leftarrow$ \Padding{$I$, $m$} \tcp{Padding.}
\end{algorithm}

\vspace{-0.5em}
\subsubsection{Data Collection.}
\label{sec:setup_and_data_collection}

\begin{figure*}[t]
\centering
\begin{minipage}[t]{0.45\textwidth}
  \vspace{0pt}
  \centering
  \includegraphics[width=\linewidth]{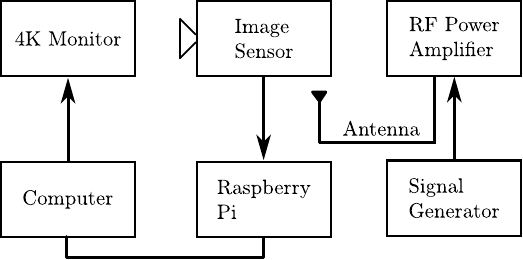}
  \captionof{figure}{The experimental setup employed for data collection.}
  \label{fig:experiment_setup}
\end{minipage}
\hfill
\begin{minipage}[t]{0.5\textwidth}
  \vspace{0pt}
  \centering
  \captionof{table}{Parameters of Attacks}
  \label{tab:exp_params}
  \footnotesize
  \setlength{\tabcolsep}{1pt}
  \begin{tabular}{@{}ll@{}}
  \hline
  \textbf{Parameter} & \textbf{Values} \\ \hline
  Freq. (MHz)          & 25, 45, 65, 85 \\
  Pow. (W)                & 2.0, 2.5, 3.0 \\
  Dur. (ms)            & 33.3, 16.7, 8.3, 2.0 \\
  Dis. (m)             & 0.02, 0.05, 0.10, 0.20 \\
  Ang. (\SI{}{\degree}) & 0, 30, 60, 90 \\
  Model      & OV5647, IMX219,378,708 \\ \hline
  \end{tabular}
\end{minipage}
\vspace{-1em}
\end{figure*}

There is no open-source ESIA image dataset, and to evaluate the simulation, we built an experimental setup to collect our own data, as depicted in Fig.~\ref{fig:experiment_setup} for the evaluation of our simulation framework.
We used a computer to display open-source images on a 4K monitor.
We randomly selected 100 photos from the COCO2017 testing dataset~\cite{lin2014microsoft}.
The image sensor is utilized to capture raw images of the content displayed on the 4K monitor. 
A cable is employed to connect the image sensor to a Raspberry Pi, which serves as the storage device for the raw images. 
The raw images are subsequently transmitted to the computer for further processing.

Regarding the attack setup, a signal generator has been employed to create the desired attacking signal. 
The generated attacking signal is then amplified using a radio-frequency (RF) power amplifier. 
An omnidirectional antenna is positioned around the cable that connects the image sensor and the Raspberry Pi, enabling wireless transmission of the attacking signal and aiming to inject and disrupt the data transmission between these components.
We choose sinusoidal waveform for the attack signals, which are commonly used in previous electromagnetic signal injection attacks that are discussed in Section~\ref{sec:background}.
The parameters of attacks are summarized in Table~\ref{tab:exp_params}.
During the data collection process, we followed a specific procedure for each picture displayed on the monitor. 
First, we captured two consecutive raw images ($I$) in the absence of any attacks. 
Next, we adjusted the parameters of attacks to result in strips within a raw image ($I^{\prime}$), and in general, an average occurrence of 3 strips, 6 strips, and 15 strips.
Empirically, we observe that stronger effective injection generally leads to more dropped rows/more strips, while strip locations depend on which specific packets are corrupted during transmission.
Our method for identifying the dropped rows is described as follows.

\subsubsection{Identifying Dropped Rows.}
\label{sec:identifying_dropped_rows}
In order to simulate the same color strips as $I^{\prime}$, it is necessary to know which rows are dropped in practical attacks, i.e., finding $D$.
This is challenging as the image sensor systems do not provide information about the dropped rows to users.
We develop the following non-trivial method to obtain it, regardless of the types of image sensors.

Recall that the two images $I^{\prime}$ and $I$ are supposed to be the same except for the dropped rows.
Although noise also causes differences, since the noise is too small to cause significant impacts, the difference between the images can be simplified as being caused by the dropped rows.
As shown in Fig.~\ref{fig:killed_rows}, we subtract $I^{\prime}$ from $I$, and keep the absolute values, i.e., ${|I - I^{\prime}|}$, and the smaller (larger) the difference, the darker (brighter) the pixel.
Before the $0^{th}$ dropped row (i.e., $x_{0}$), rows in $I^{\prime}$ and rows in $I$ are the same, and their difference is zero, leading to dark pixels in ${|I - I^{\prime}|}$.
When the $0^{th}$ dropped row occurs until the first ($x_{1}$), rows between the two images are different, leading to bright pixels in ${|I - I^{\prime}|}$. 
After the first but before the second, the color transformation of the rows returns to normal; although the difference between pixel values in the two images is not zero, they are still pretty small, causing dark pixels in ${|I - I^{\prime}|}$.
So on, if we compute and visualize the difference between $I^{\prime}$ and $I$, it can be found that the bright regions correspond to the color strips in the reconstructed image.

Based on such an observation, we can use edge detection methods~\cite{2020SciPy-NMeth} to identify the positions of the color strips (i.e., $x_{i}^{\prime}$) in $I^{\prime}$.
Because the dropped rows will shift the following rows up, it is essential to point out that the positions of the edges are not where the rows are dropped.
The relationship between the edges and their corresponding dropped rows can be derived from Equation~\ref{eq:position_of_strips}:
\begin{equation}
    x_{i} = 
    \begin{cases}
        x_{i}^{\prime}, & \text{if}~i = 0, \\
        x_{i}^{\prime}  + 2 \times \lfloor \frac{i + 1}{2} \rfloor, & \text{otherwise.}
    \end{cases}
\end{equation}
Therefore, we can obtain the dropped rows ($D$) in each image under attack.
Together with the raw image ($I$), the simulation method can be applied to generate synthetic images ($I^{s}$).

\section{Evaluation of Synthetic Adversarial Images}
\label{sec:evaluation}

As discussed in Section~\ref{sec:introduction}, it is crucial to verify the equivalence of our synthetic adversarial images to real-world attacks.
To accomplish this, our system model provides effective guidance through the following three aspects. (1) $I^{\prime}$ v.s. $I^{s}$: Examining the similarity between raw images; (2)$R(I^{\prime})$ v.s. $R(I^{s})$: Evaluating the similarity between reconstructed images; (3) $C(R(I^{\prime}))$ v.s. $C(R(I^{s}))$: Determining whether the reconstructed images have similar impacts on the computer vision algorithms.



\subsection{Metrics}

For the first two evaluations, we have chosen the Structural Similarity Index Measure (SSIM) as the metric, which is widely used in image processing and computer vision to quantify the similarity between two images~\cite{wang2004image}.
The SSIM ranges from 0 to 1, where a value closer to 1 indicates a higher similarity between the images. 

The last evaluation is about assessing the performance of computer vision tasks.
We selected object detection for evaluation later because its vulnerability has been demonstrated in previous studies~\cite{jiang23glitchhiker,zhang2024esia}, as well as a similar task, instance segmentation; these tasks are fundamental to other more advanced computer vision tasks~\cite{manakitsa2024review}.
The common metric for such tasks is Mean Average Precision (mAP)~\cite{padilla2020survey}.
The mAP score ranges from 0 to 1, with a higher value indicating a better detection performance.
Note that there are different variants of mAP that will be used hereafter: mAP50, mAP75, and mAP50:95 measure the mean average precision at an intersection over union (IoU) threshold of 0.5, 0.75, and across IoU thresholds ranging from 0.5 to 0.95, respectively.

\subsection{Structural Similarity between Raw Images}

\begin{figure}[t]
\centering
\includegraphics[width=0.5\textwidth]{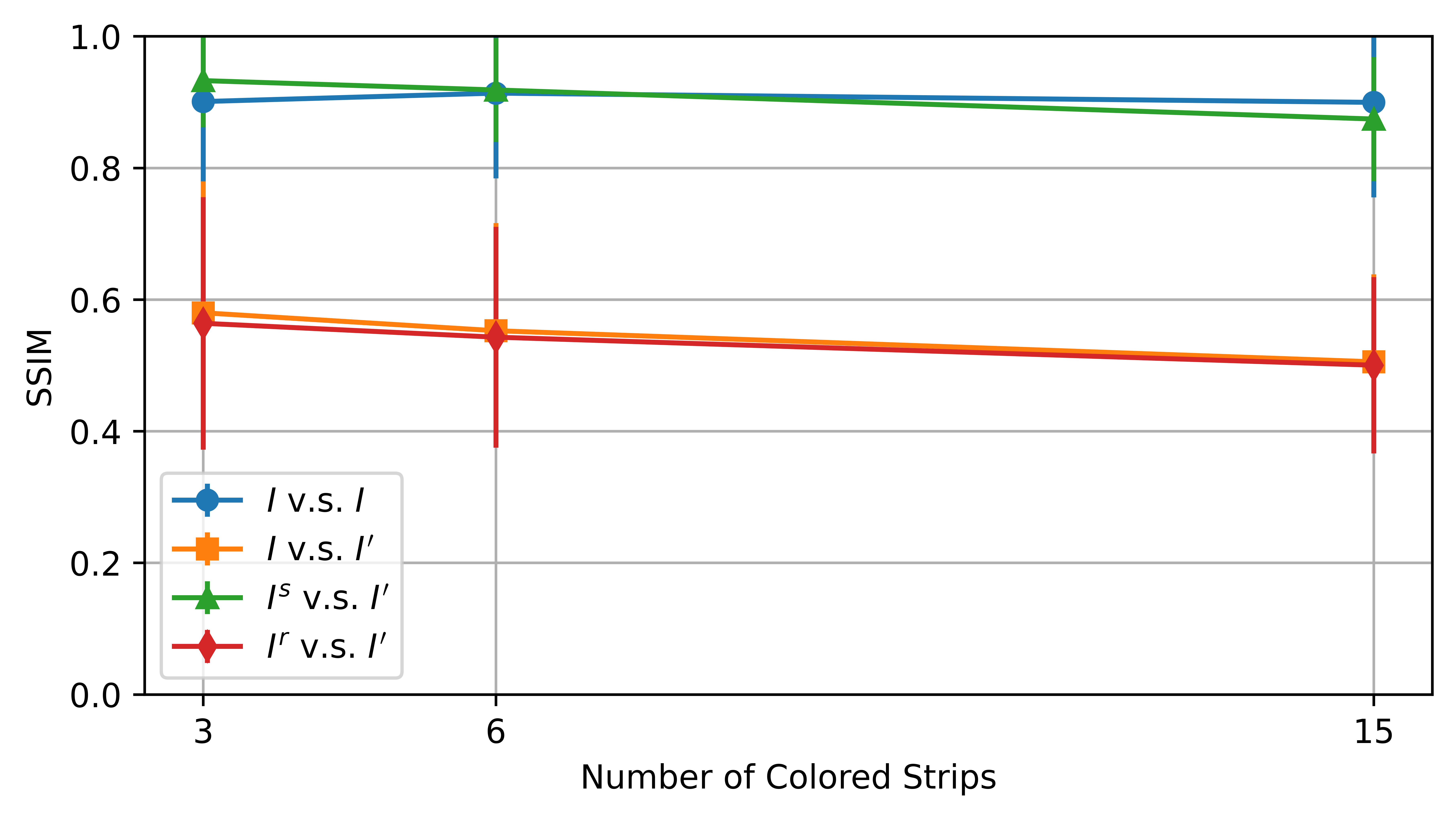}
\caption{SSIM between different raw images. The synthetic adversarial images are highly similar to real attack images.}
\label{fig:ssim_raw}
\vspace{-1em}
\end{figure}

\begin{table*}[t]
\centering
\caption{SSIM under different attack parameters with the default configuration of \SI{85}{\mega\hertz}, \SI{3.0}{\watt}, \SI{33.3}{\milli\second}, \SI{0.02}{\meter}, \SI{0}{\degree}, and OV5647.}
\label{tab:ssim_under_diff_params}
\footnotesize
\resizebox{\textwidth}{!}{%
\begin{tabular}{ll llll@{\hskip 12pt} lll@{\hskip 12pt} llll@{\hskip 12pt} lll@{\hskip 12pt} lll@{\hskip 12pt} lll}
\hline
SSIM & & \multicolumn{4}{l@{\hskip 12pt}}{Freq. (MHz)} & \multicolumn{3}{l@{\hskip 12pt}}{Pow. (W)} & \multicolumn{4}{l@{\hskip 12pt}}{Dur. (ms)} & \multicolumn{3}{l@{\hskip 12pt}}{Dist. (m)} & \multicolumn{3}{l@{\hskip 12pt}}{Ang. (\si{\degree})} & \multicolumn{3}{l}{Sensor (IMX)} \\
($\cdot,\cdot$)& & 25 & 45 & 65 & 85 & 2 & 2.5 & 3 & 2 & 8.3 & 16.7 & 33.3 & 0.05 & 0.1 & 0.2 & 30 & 60 & 90 & 219 & 378 & 708 \\
\hline
\multirow{2}{*}{($I^{\prime}$, $I^{s}$)} & $\mu$ & 0.96 & 0.94 & 0.95 & 0.84 & 0.96 & 0.93 & 0.89 & 0.96 & 0.96 & 0.94 & 0.91 & 0.94 & 0.95 & 0.91 & 0.95 & 0.94 & 0.91 & 0.84 & 0.95 & 0.88 \\
& $\sigma$ & 0.01 & 0.06 & 0.08 & 0.09 & 0.02 & 0.07 & 0.10 & 0.02 & 0.02 & 0.03 & 0.08 & 0.04 & 0.03 & 0.11 & 0.03 & 0.05 & 0.06 & 0.03 & 0.04 & 0.13 \\
\multirow{2}{*}{($I^{\prime}$, $I^{r}$)} & $\mu$ & 0.57 & 0.46 & 0.43 & 0.46 & 0.52 & 0.49 & 0.45 & 0.33 & 0.54 & 0.63 & 0.50 & 0.55 & 0.61 & 0.33 & 0.42 & 0.40 & 0.45 & 0.32 & 0.49 & 0.75 \\
& $\sigma$ & 0.27 & 0.21 & 0.29 & 0.27 & 0.27 & 0.31 & 0.21 & 0.32 & 0.16 & 0.25 & 0.10 & 0.28 & 0.28 & 0.19 & 0.20 & 0.21 & 0.20 & 0.38 & 0.24 & 0.18 \\
\hline
\end{tabular}%
}
\vspace{-1em}
\end{table*}

We first measured SSIM between paired non-attack raw images and obtained an average of about 0.9, which serves as the upper-bound reference value in our dataset. In contrast, the average SSIM between a real attack image \(I'\) and its corresponding clean image \(I\) drops to about 0.5, representing the lower-bound reference. Comparing the real attack image \(I'\) with our synthetic image \(I^{s}\), we again observe an average SSIM of about 0.9, indicating strong raw-level similarity. By comparison, a random row-drop simulation \(I^{r}\) achieves only about 0.5 SSIM, showing that our method captures dropped rows much more accurately than a random strategy. Under varied experimental conditions in Table~\ref{tab:exp_params}, as shown in Table~\ref{tab:ssim_under_diff_params}, \(I^{s}\) consistently maintains high fidelity to \(I'\), with mean \(\mathrm{SSIM}(I', I^{s})\) above 0.84, whereas \(\mathrm{SSIM}(I', I^{r})\) can be as low as 0.33. Overall, our method improves SSIM by roughly 0.4 over the random baseline.
Considering 95\% confidence intervals for the performance differences, these intervals are narrow, providing stronger evidence that the synthetic adversarial images closely approximate the effects of real attacks.

\subsection{Structural Similarity between Reconstructed Images}
\label{sec:ssim_between_reconstructed_images}

Image reconstruction converts a raw image into a full-color image and is affected by factors such as demosaicing, white balancing, and noise reduction. Since many reconstruction functions exist, our goal is not to evaluate them exhaustively, but to examine the similarity between the reconstructed images \(R(I^{\prime})\) and \(R(I^{s})\). To approximate the reconstruction function \(R(\cdot)\), we adopt the widely used Adaptive Homogeneity-Directed (AHD) demosaicing algorithm~\cite{hirakawa2005adaptive}.

\begin{figure}[t]
\centering
\includegraphics[width=0.5\textwidth]{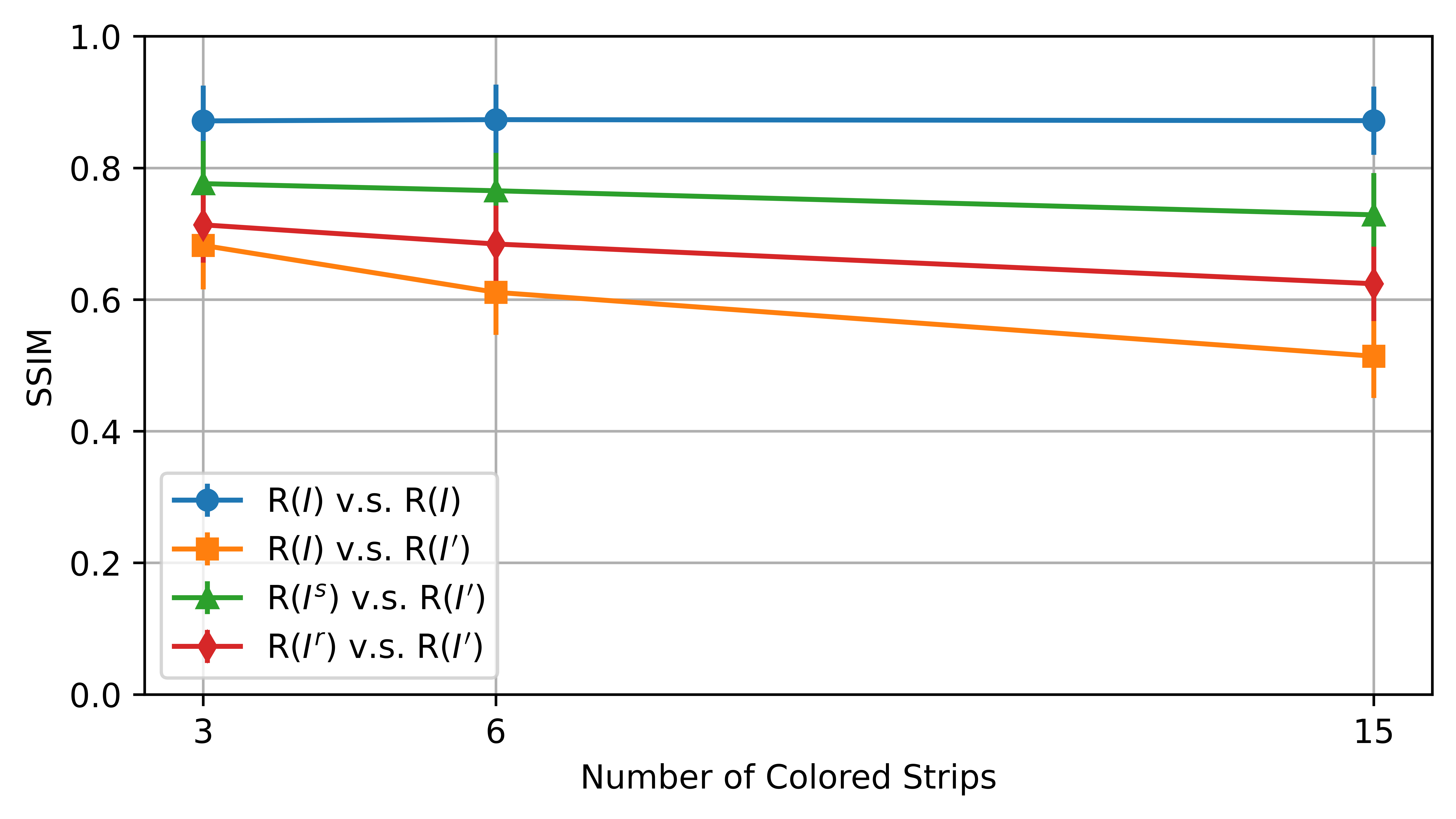}
\caption{SSIM between different reconstructed images. The synthetic adversarial images are fairly similar to real attack images.}
\label{fig:ssim_reconstructed}
\vspace{-1em}
\end{figure}

Consistent with the previous subsection, we have considered four different reconstructed images: $R(I)$, $R(I^{\prime})$, $R(I^{s})$, and $R(I^{r})$. We then calculated the SSIM for the following four comparison combinations, as illustrated in Fig.~\ref{fig:ssim_reconstructed}.
As we can see, the highest similarity has been achieved by $R(I)$~v.s.~$R(I)$  with average SSIM of $0.87$, followed by $R(I^{s})$~v.s.~$R(I^{\prime})$, and then $R(I^{r})$~v.s.~$R(I^{\prime})$, and finally $R(I)$~v.s.~$R(I^{\prime})$ with the lowest similarity. 
From the results, the images reconstructed by using the real attack images and the images reconstructed by synthetic adversarial images still displayed a similarity as high as $0.80$, statistically significantly higher than the random simulation of $0.67$.
It shows that our simulation method did manage to synthesize the attack with a high degree of resemblance.

\subsection{Performance Similarity}
\label{sec:performance_similarity}
Our focus lies on assessing the performance of two computer vision tasks: object detection and instance segmentation, as mentioned previously. 
The objective is to evaluate the impact of both real and synthetic adversarial images on these tasks, and demonstrate their comparable effects on different algorithms/models.

We began by employing the state-of-the-art models that have been benchmarked on the COCO dataset~\cite{lin2014microsoft} to generate preliminary ground-truth labels for the tasks of object detection and image segmentation.
Regarding object detection, we chose Co-DETR~\cite{zong2023detrs} which stands out as one of the state-of-the-art models, while for instance segmentation, we chose Mask2Former~\cite{cheng2021mask2former}.
These models provide an initial set of high-quality labels. 
Subsequently, we conducted a thorough manual review of these automatically generated labels. 
During this review, we adjusted and refined each label, which included correcting misclassified labels and fine-tuning the boundaries of the bounding boxes.
This meticulous process ensures the accuracy of the labels used in our evaluation.



\subsubsection{Object Detection.}
\begin{figure}[t]
\centering
\includegraphics[width=0.78\textwidth]{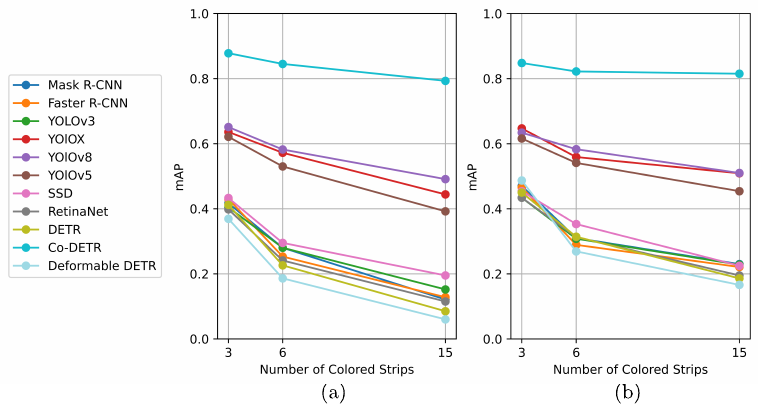}
\caption{Similar performance (mAP50) between (a) real attack images and (b) synthetic adversarial images across different object detection models.}
\label{fig:object_detection_mAP}
\vspace{-1em}
\end{figure}

We selected representative object detection models spanning diverse architectures, including Mask R-CNN~\cite{he2017mask}, Faster R-CNN~\cite{ren2015faster}, YOLO~\cite{redmon2016you} variants (v3, v5, X, and v8), SSD~\cite{liu2016ssd}, RetinaNet~\cite{lin2017focal}, DETR~\cite{carion2020end}, Co-DETR~\cite{zong2023detrs}, and Deformable DETR~\cite{zhu2021deformable}. We evaluated these models on both real attack images and synthetic adversarial images using mAP. As shown in Figure~\ref{fig:object_detection_mAP}, mAP consistently decreases as the number of color strips increases in both settings. To compare the two statistically, we aggregated results across models and performed Mann-Whitney U tests on mAP50, mAP75, and mAP50:90 at a \(5\%\) significance level. Since all p-values in Table~\ref{tab:mwu-test} exceed \(5\%\), we fail to reject the null hypothesis, supporting that synthetic adversarial images effectively replicate the impact of real attacks.

\begin{table}[t]
\footnotesize
\centering
\caption{p-values of Mann-Whitney U rank test.}
\label{tab:mwu-test}
\setlength{\tabcolsep}{8pt}
\resizebox{0.6\columnwidth}{!}{%
\begin{tabular}{@{}l ccc ccc@{}}
\toprule
 & \multicolumn{3}{c}{\textbf{Object Detection}} & \multicolumn{3}{c}{\textbf{Segmentation}} \\
\cmidrule(lr){2-4} \cmidrule(l){5-7}
\textbf{\# of strips} & \textbf{3} & \textbf{6} & \textbf{15} & \textbf{3} & \textbf{6} & \textbf{15} \\
\midrule
mAP50    & 0.148 & 0.189 & 0.094 & 0.238 & 0.166 & 0.182 \\
mAP75    & 0.767 & 0.149 & 0.149 & 0.369 & 0.137 & 0.166 \\
mAP50:95 & 0.293 & 0.101 & 0.108 & 0.305 & 0.174 & 0.174 \\
\bottomrule
\end{tabular}%
}
\vspace{-1em}
\end{table}

\begin{figure}[t]
\centering
\includegraphics[width=0.78\textwidth]{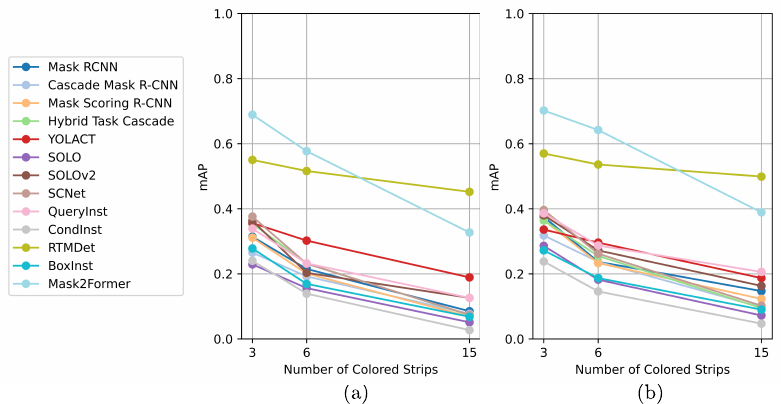}
\caption{Similar performance (mAP50) between (a) real attack images and (b) synthetic adversarial images across different instance segmentation models.}
\label{fig:segmentation_mAP}
\vspace{-1em}
\end{figure}
\subsubsection{Instance Segmentation.}
We selected representative instance segmentation models, including Mask R-CNN~\cite{he2017mask}, Cascade Mask R-CNN~\cite{cai2019cascade}, Mask Scoring R-CNN~\cite{huang2019mask}, Hybrid Task Cascade~\cite{chen2019hybrid}, RTMDet~\cite{lyu2022rtmdet}, CondInst~\cite{tian2020conditional}, BoxInst~\cite{tian2021boxinst}, Mask2Former~\cite{cheng2021mask2former}, SCNet~\cite{vu2021scnet}, QueryInst~\cite{fang2021instances}, and SOLO~\cite{wang2020solo,wang2020solov2}, to evaluate attack impact. These models cover diverse design choices and strong segmentation performance. We then tested them on both real attack images and synthetic adversarial images and measured mAP. As shown in Fig.~\ref{fig:segmentation_mAP}, mAP consistently declines as the number of color strips increases for both image types. To compare their effects statistically, we aggregated results across models and conducted Mann-Whitney U tests on mAP50, mAP75, and mAP50:95. Since all p-values in Table~\ref{tab:mwu-test} are above the 
5\%
significance level, we do not detect a statistically significant difference between the effects of real and synthetic adversarial images under the Mann–Whitney U test.

\section{Vulnerability Assessment and Robustness Enhancement}
\label{sec:case_study}
\subsection{Vulnerability Assessment of Different Computer Vision Tasks}
\label{sec:case_study_different_cv_tasks}
In this section, we will demonstrate how to use our simulation method to explore the impacts of electromagnetic signal injection attacks on different models. 
Note that we will use public datasets, but images in these datasets are not in the raw format.
Therefore, we applied the state-of-the-art conversion tool~\cite{zamir2020cycleisp} to obtain the raw format from RGB images, and then applied our simulation method to generate synthetic adversarial images, with the number of strips ranging from 1 to 20, where their positions are random.
Note that the conversion model reported an SSIM above 0.99 between the reconverted RGB images and the original ground truth.

\subsubsection{Models and Datasets.}

In addition to object detection and instance segmentation, we considered three other vision tasks: (1) image classification, (2) face recognition, and (3) depth estimation. For image classification, we selected ResNet50~\cite{he2016deep}, Inception v3~\cite{szegedy2016rethinking}, and ViT~\cite{dosovitskiy2020image}, and evaluated accuracy on the ILSVRC 2012 validation set~\cite{ILSVRC15}. 
For face recognition, we used FaceNet~\cite{schroff2015facenet}, ArcFace~\cite{deng2018arcface}, and AdaFace~\cite{kim2022adaface}, and measured 1:1 verification accuracy (TPR@FPR = \(10^{-4}\)) on the IJB-C dataset~\cite{maze2018iarpa}. 
For depth estimation, we selected BTS~\cite{lee2019big}, DPT-Hybrid~\cite{Ranftl2021}, and Depth Anything~\cite{depthanything}, and evaluated performance using SSIM on the KITTI dataset~\cite{Geiger2012CVPR}.
These benchmark datasets provide standardized and widely adopted testbeds for evaluating model performance under adversarial conditions. They also enable efficient experimentation across multiple tasks without the complexity of setting up real-world attacks, allowing us to study attack effects under diverse scenarios.

\begin{table*}[t]
\footnotesize
\centering
\caption{Vulnerability Assessment of Different Computer Vision Tasks to Attacks}
\label{tab:susceptibility_of_different_cv_tasks}
\resizebox{0.85\columnwidth}{!}{%
\begin{tabular}{llllll}
\hline
\textbf{Tasks} & \textbf{Metric} & \textbf{Models} & \textbf{No Attack} & \textbf{Attack} & \textbf{Degradation} \\ \hline
Image Classification & Accuracy & \begin{tabular}[c]{@{}l@{}}ResNet50~\cite{he2016deep}\\ Inception v3~\cite{szegedy2016rethinking}\\ ViT~\cite{dosovitskiy2020image}\end{tabular} & \begin{tabular}[c]{@{}l@{}}0.71 \\ 0.68 \\ 0.70\end{tabular} & \begin{tabular}[c]{@{}l@{}}0.49\\ 0.57 \\ 0.55\end{tabular} & \begin{tabular}[c]{@{}l@{}}-31\%\\ -16\%\\ -21\%\end{tabular} \\ \hline
Face Recognition & Accuracy & \begin{tabular}[c]{@{}l@{}}FaceNet~\cite{schroff2015facenet}\\ ArcFace~\cite{deng2018arcface}\\ AdaFace~\cite{kim2022adaface}\end{tabular} & \begin{tabular}[c]{@{}l@{}}0.45\\ 0.84 \\ 0.87\end{tabular} & \begin{tabular}[c]{@{}l@{}}0.35\\ 0.77 \\ 0.81\end{tabular} & \begin{tabular}[c]{@{}l@{}}-22\%\\ -9\%\\ -6\%\end{tabular} \\ \hline
Depth Estimation & SSIM & \begin{tabular}[c]{@{}l@{}}BTS~\cite{lee2019big}\\ DPT~\cite{Ranftl2021}\\ Depth Anything~\cite{depthanything}\end{tabular} & \begin{tabular}[c]{@{}l@{}}0.93 \\ 0.93\\ 0.94\end{tabular} & \begin{tabular}[c]{@{}l@{}}0.91\\ 0.92\\ 0.94\end{tabular} & \begin{tabular}[c]{@{}l@{}}-1\%\\ -1\%\\ 0\%\end{tabular} \\ \hline
Object Detection & mAP50 & \begin{tabular}[c]{@{}l@{}}YOLOv8~\cite{redmon2016you}\\ Mask R-CNN~\cite{he2017mask}\\ DETR~\cite{carion2020end}\end{tabular} & \begin{tabular}[c]{@{}l@{}}0.70 \\ 0.69\\ 0.63\end{tabular} & \begin{tabular}[c]{@{}l@{}}0.57\\ 0.34\\ 0.31\end{tabular} & \begin{tabular}[c]{@{}l@{}}-19\%\\ -51\%\\ -51\%\end{tabular} \\ \hline
Instance Segmentation & mAP50 & \begin{tabular}[c]{@{}l@{}}SOLOv2~\cite{wang2020solov2}\\ Mask R-CNN~\cite{he2017mask}\\ RTMDet~\cite{lyu2022rtmdet}\end{tabular} & \begin{tabular}[c]{@{}l@{}}0.53\\ 0.53\\ 0.60\end{tabular} & \begin{tabular}[c]{@{}l@{}}0.27\\ 0.25 \\ 0.54\end{tabular} & \begin{tabular}[c]{@{}l@{}}-49\%\\ -53\%\\ -10\%\end{tabular} \\ \hline
\end{tabular}
}
\vspace{-1em}
\end{table*}

\subsubsection{Results and Analysis.}
The results are summarized in Table~\ref{tab:susceptibility_of_different_cv_tasks}, where ``Degradation'' denotes the percentage performance drop under attack relative to the no-attack condition, hereafter. Overall, nearly all models suffer performance degradation under electromagnetic signal injection attacks, with object detection and instance segmentation showing the greatest vulnerability. These tasks experience drops of up to \(53\%\), whereas image classification degrades more moderately, by about \(16\%\) to \(31\%\), with Inception v3 being the most resilient.
This gap is likely due to task complexity. Unlike image classification, object detection, and instance segmentation require not only recognition but also accurate localization and mask prediction, giving attackers more opportunities to disrupt performance. By contrast, face recognition and depth estimation are more robust. Face recognition models show smaller accuracy drops, with AdaFace degrading by only \(6\%\). Depth estimation appears especially robust, although this is partly because evaluation masks predictions using ground truth, which may underestimate attack impact in real deployment.
These results reveal substantial differences in vulnerability across vision tasks and motivate further study of the causes of such degradation to support the design of effective defenses.

\subsection{Robustness Enhancement by Adversarial Training}
\label{sec:case_study_funetuning}

\begin{table*}[t]
\caption{Performance of Object Detection Models before/after Fine-tuning.}
\label{tab:fine_tuning}
\centering
\footnotesize
\resizebox{0.85\columnwidth}{!}{%
\begin{tabular}{llllll}
\hline
\textbf{Model} & \textbf{Train} & \textbf{Validation} & \textbf{\begin{tabular}[c]{@{}l@{}}mAP50\\ (Degradation)\end{tabular}} & \textbf{\begin{tabular}[c]{@{}l@{}}mAP75\\ (Degradation)\end{tabular}} & \textbf{\begin{tabular}[c]{@{}l@{}}mAP50:95\\ (Degradation)\end{tabular}} \\ \hline
\multirow{4}{*}{DETR~\cite{carion2020end}} & - & No Attack & 0.51 & 0.33 & 0.35 \\
 & - & Attack & 0.25 (-50\%) & 0.16 (-52\%) & 0.16 (-55\%) \\
 & No Attack & Attack & 0.21 (-59\%) & 0.14 (-59\%) & 0.13 (-63\%) \\
 & Attack & Attack & 0.36 (-30\%) & 0.24 (-29\%) & 0.22 (-37\%) \\ \hline
\multirow{4}{*}{YOLOX~\cite{redmon2016you}} & - & No Attack & 0.63 & 0.47 & 0.46 \\
 & - & Attack & 0.52 (-17\%) & 0.37 (-22\%) & 0.37 (-20\%) \\
 & No Attack & Attack & 0.46 (-27\%) & 0.36 (-23\%) & 0.33 (-27\%) \\
 & Attack & Attack & 0.55 (-13\%) & 0.43 (-9\%) & 0.40 (-12\%) \\ \hline
\multirow{4}{*}{Faster R-CNN~\cite{ren2015faster}} & - & No Attack & 0.55 & 0.39 & 0.36 \\
 & - & Attack & 0.29 (-47\%) & 0.18 (-54\%) & 0.18 (-50\%) \\
 & No Attack & Attack & 0.32 (-42\%) & 0.21 (-47\%) & 0.20 (-44\%) \\
 & Attack & Attack & 0.34 (-37\%) & 0.22 (-43\%) & 0.22 (-39\%) \\ \hline
\end{tabular}
}
\vspace{-1em}
\end{table*}

We perform adversarial training by fine-tuning models on attack images to enhance their robustness against the attacks.
In this section, we will demonstrate its effectiveness using object detection tasks as an example.
Note that we are not proposing a new adversarial training framework; rather, we show how the proposed synthetic adversarial images can be directly leveraged within existing training pipelines to improve model robustness.

\subsubsection{Models and Datasets.}

For adversarial training evaluation, we selected three representative object detection models: DETR~\cite{carion2020end}, YOLOX~\cite{redmon2016you}, and Faster R-CNN~\cite{ren2015faster}, covering transformer-based, one-stage, and two-stage architectures, respectively. We randomly sampled 1000 images from the COCO validation set and split them into training and validation subsets with a 7:3 ratio while preserving category coverage.

To assess the benefit of fine-tuning on attack images, we constructed two training sets: one with original clean images and the other with synthetic adversarial images. All models were fine-tuned for 150 epochs on each set and then evaluated on the validation set, enabling comparison between standard fine-tuning and attack-aware fine-tuning.

For synthetic adversarial images, annotations must be adjusted to account for the row-shifting effect described in Section~\ref{sec:simulation_method}. Given an original bounding box \([center_x, center_y, height, width]\), let \(n_1\) denote the number of lost lines before \(center_y\), and \(n_2\) the number lost of rows within the box. The adjusted box becomes \([center_x, center_y - n_1, height - n_2, width]\), ensuring proper alignment with the modified image content.

\subsubsection{Results and Analysis.}

The results are shown in Table~\ref{tab:fine_tuning}. Before fine-tuning, all models suffer substantial performance drops on images with synthetic adversarial patterns. For example, DETR's mAP50:95 decreases from 0.35 to 0.16, corresponding to a \(55\%\) degradation.
Fine-tuning only on clean images provides little benefit against attacks. In most cases, such as DETR and YOLOX, the performance on adversarial images remains largely unchanged, suggesting that standard fine-tuning does not generalize well to adversarial settings. By contrast, fine-tuning with synthetic adversarial images significantly improves robustness across all models. DETR's mAP50:95 rises from 0.16 to 0.22, reducing degradation from \(55\%\) to \(37\%\). YOLOX, which has the strongest clean-image baseline, further improves to 0.40 mAP50:95 under attack, retaining \(88\% (= 1-12\%)\) of its original performance; for mAP75, it retains up to \(91\%\). Faster R-CNN also improves from 0.18 to 0.22 in mAP50:95, reducing degradation from \(50\%\) to \(39\%\).
Although adversarial training is a well-known defense, these results show that even a simple form based on synthetic adversarial images can improve robustness. 

\section{Discussion, Limitation, and Future Work}
\label{sec:discussion}



Manual image collection, which requires configuring the attack signal before capturing each photo, takes an average of 96 seconds per image.
However, our simulation is faster. 
Using an Apple M1 system, the simulation method generates an attacked RGB image from a raw image in only 0.49 seconds. 
This speed improvement, with simulation being about 196 times faster than manual collection (96 seconds / 0.49 seconds $\approx$ 195.9), improves how efficiently we can create attack images. 
The ability to quickly make attack images allows for faster research and development, making it easier to test different attacks. 

Although the proposed framework can effectively reproduce row-drop-based color strip artifacts, it depends on knowing which rows are dropped, and therefore should be viewed as a simulator of the image-level consequences of ESIA rather than a complete physical emulator of the attack process. 
The current method does not establish a direct mapping from physical RF parameters (e.g., frequency, power, and waveforms) to the number, position, or distribution of dropped rows, leaving a gap for future work. 
In addition, the reconstruction pipeline is simplified: evaluations rely mainly on AHD demosaicing and do not yet account for the diversity of real Image Signal Processing (ISP) pipelines, including white balance, denoising, gamma correction, compression, etc., all of which are usually proprietary to different image sensor manufacturers. 
Finally, the robustness study is preliminary. In our adversarial training experiments, both training and testing are conducted on synthetic attacks. Whether models enhanced on synthetic data transfer to physically captured ESIA images remains to be further investigated in future work.

\section{Conclusion}
\label{sec:conclusion}

In this work, we tackle the practical barriers that have hindered research into electromagnetic signal injection attacks (ESIA) on image sensor systems, particularly the difficulty of obtaining adversarial data without specialized hardware. 
To bridge this gap, we introduce a simulation framework for generating adversarial patterns in images, grounded in a theoretical model of ESIA. This framework enables reproducible evaluations of downstream AI under attack, eliminating the need for complex physical setups.
Through extensive experiments, we demonstrate that our simulation approximates real-world ESIA patterns. 
Using the synthetic adversarial data, we show that a wide range of computer vision applications are highly susceptible to such attacks, highlighting an urgent need for more robust defenses.
Moreover, our preliminary study on adversarial training reveals its potential effectiveness in enhancing model robustness.

\textbf{Acknowledgment.} This work was in part supported by Hong Kong RGC Project (No. PolyU15227825), and The Hong Kong Polytechnic University under grants P0059492, P0062002, and P0048514.


\bibliographystyle{splncs04} 
\bibliography{main}

\end{document}